\documentclass[ALICE,manyauthors]{cernphprep}
\pdfoutput=1
\usepackage[%
            final,
            color,
            eps 
            ]{style}
\begin{document}
\currentpdfbookmark{Front}{front}
\begin{titlepage}
  \PHyear{2016} 
  \PHnumber{327} 
  \PHdate{30\textsuperscript{th}{} December{}}
  \title{Centrality dependence of the pseudorapidity density
    distribution for charged particles in \PbPb{} collisions at
    $\usNN*{PbPb}{5023}$}
  \ShortTitle{Charged-particle pseudorapidity density in \PbPb{} at
    $\usNN{PbPb}{5023}$}%
  \Collaboration{ALICE Collaboration\thanks{See
      Appendix~\ref{app:collab} for the list of collaboration
      members}}%
  \ShortAuthor{ALICE Collaboration}%
  \abstract{ 
We present the charged-particle pseudorapidity density in \PbPb{}
collisions at $\usNN{PbPb}{5023}$ in centrality classes measured by
ALICE.  The measurement covers a wide pseudorapidity range from $-3.5$
to $5$, which is sufficient for reliable estimates of the total number
of charged particles produced in the collisions.  For the most central
(\centRange{0}{5}) collisions we find $21\,400\pm 1\,300$, while for
the most peripheral (\centRange{80}{90}) we find $230\pm 38$.  This
corresponds to an increase of $\per{(27\pm4)}$ over the results at
$\usNN{PbPb}{2760}$ previously reported by ALICE.  The energy
dependence of the total number of charged particles produced in
heavy-ion collisions is found to obey a modified power-law like
behaviour.  The charged-particle pseudorapidity density of the most
central collisions is compared to model calculations --- none of which
fully describes the measured distribution. We also present an estimate
of the rapidity density of charged particles. The width of that
distribution is found to exhibit a remarkable proportionality to the
beam rapidity, independent of the collision energy from the top SPS to
LHC energies.


} \MarkSvnVersion{}
\end{titlepage}
\addtocounter{page}{-1}

\section{Introduction}
\label{sec:intro}

In ultra-relativistic heavy-ion collisions a dense and hot phase of
nuclear matter is
created~\cite{Arsene:2004fa,Back:2004je,Adams:2005dq,Adcox:2004mh}.
This phase of QCD matter is considered to be a plasma of strongly
interacting quarks and gluons and is therefore labelled the
sQGP~\cite{Nagle:2006cj}.  The multiplicity of primary, charged
particles produced in heavy-ion collisions is a key observable to
characterise the properties of the matter created in these collisions
\cite{Armesto:2009ug}. The study of the primary charged-particle
pseudorapidity density ($\dndeta$) over a wide pseudorapidity ($\eta$)
range and its dependence on colliding system, centre-of-mass energy,
and collision geometry is important to understand the relative
contributions to particle production from hard scatterings and soft
processes, and may provide insight into the partonic structure of the
interacting nuclei.

\belowpdfbookmark{Previous results, and now}{intro:previous}%
We have previously reported measurements on primary charged-particle
pseudorapidity densities over a wide pseudorapidity range in Pb--Pb
collisions at the centre-of-mass energy per nucleon pair
$\usNN{PbPb}{2760}$~\cite{Adam:2015kda}.  In this Letter, we study
these distributions in the pseudorapidity interval from $-3.5$ to $5$
at a collision energy of $\usNN{PbPb}{5023}$ as a function of the
centrality. Pseudorapidity is defined as
$\eta\equiv-\log(\tan(\vartheta/2))$, where $\vartheta$ is the angle
between the charged-particle trajectory and the beam axis
($z$--axis). Nuclei are extended objects, and their collisions can be
characterised by centrality --- the experimental proxy for the
un-measurable distance between the centres of the colliding nuclei
(impact parameter).
A primary particle is a particle with a mean proper lifetime $\tau$
larger than $\unit[1]{cm\kern-.05em/\kern-.1em c}$, which is either a)
produced directly in the interaction, or b) from decays of particles
with $\tau$ smaller than $\unit[1]{cm\kern-.05em/\kern-.1em c}$,
restricted to decay chains leading to the
interaction~\cite{ALICE-PUBLIC-2017-005}.
In this Letter, all quantities reported are for
primary charged particles, though we will omit ``primary'' for
brevity.

\belowpdfbookmark{Total number of charged particles}{intro:TotalNch}%
With the large pseudorapidity coverage available in ALICE, we can
reliably estimate, for all centrality classes, the total number of
charged particles produced in the collisions.  We therefore also
present the first measurement of the total charged-particle
multiplicity in Pb--Pb collisions at $\usNN{PbPb}{5023}$ as a function
of the number of nucleons participating in the collisions
($N_{\text{part}}$).

\belowpdfbookmark{Charged-particle rapidity density}{intro:dNdy}%
Finally, we transform the measured $\dndeta$ distribution for the
\per{5} most central collisions into charged-particle rapidity density
($\dndy$), and we examine the centre-of-mass energy dependence of the
width of that distribution.  The rapidity ($y$) of a particle with
energy $E$ and momentum component $p_z$ along the beam axis is defined
as $y\equiv\frac12\log\left([E+p_{\text{z}}]/[E-p_{\text{z}}]\right)$.
The comparison of the width of the $\dndy$ at different collision
energies provides an insight into the constraints on the overall
production mechanism of charged particles.


 
\section{Experimental setup}
\label{sec:exper}

A detailed description of ALICE and its performance can be found
elsewhere~\cite{Aamodt:2008zz,Abelev:2014ffa}. In the following, we
briefly describe the detectors relevant to this analysis.

\belowpdfbookmark{SPD}{exper:spd}%
The Silicon Pixel Detector (SPD), the innermost part of the Inner
Tracking System (ITS), consists of two cylindrical layers of hybrid
silicon pixel assemblies covering $|\eta|<2$ and $|\eta|<1.4$ for the
inner and outer layers, respectively.  Combinations of hits on each of
the two layers consistent with tracks originating from the interaction
point form \emph{tracklets}.

\belowpdfbookmark{FMD}{exper:fmd}%
The Forward Multiplicity Detector (FMD) is a silicon strip detector
which, records the energy deposited by particles traversing the
it.  The detector covers the pseudorapidity regions
$-3.5<\eta<-1.8$ and $1.8<\eta<5$, and has almost full coverage in
azimuth ($\varphi$), and high granularity in the radial ($\eta$)
direction.

\belowpdfbookmark{V0}{exper:v0}%
The third detector system used in this analysis is the V0.  It
consists of two sub-detectors: \mbox{V0\nolinebreak-\nolinebreak A}
and \mbox{V0\nolinebreak-\nolinebreak C} covering the pseudorapidity
regions $2.8<\eta<5.1$ and $-3.7<\eta<-1.7$, respectively, each made
up of scintillator tiles with a timing resolution $<\unit[1]{ns}$.
The fast signals from either of \mbox{V0\nolinebreak-\nolinebreak A}
or \mbox{V0\nolinebreak-\nolinebreak C} are combined in a programmable
logic to form a trigger signal and to reject background events.
Furthermore, the combined pulse height signal of both sub-detectors
forms the basis for the classification of events into different
centrality classes~\cite{Abelev:2013qoq}.

\belowpdfbookmark{ZDC}{exper:zdc}
The Zero--Degree Calorimeter (ZDC) measures the energy of spectator
(non--interacting) nucleons with two components: one measures protons
and the other measures neutrons. The ZDC is located at about
\unit[112.5]{m} from the interaction point on both sides of the
experiment~\cite{Aamodt:2008zz}.  The ZDC also provides timing
information used to select collisions in the off-line data processing.


\section{Data sample and analysis method}
\label{sec:data}
 
\belowpdfbookmark{Data set}{data:dataset}%
The results presented here are based on data collected by ALICE in
2015 during the \PbPb{} collision run of the LHC at
$\usNN{PbPb}{5023}$.  About $100\,000$ events with a minimum bias
trigger requirement \cite{Aamodt:2010cz} were analysed in the
centrality range from $\per{0}$ to $\per{90}$.  The minimum bias
trigger for Pb--Pb collisions in ALICE, which defines the so-called
visible cross-section, is defined as a coincidence between the A
($z>0$) and C ($z<0$) sides of the V0 detector.

\belowpdfbookmark{Event and centrality selection}{data:selection}%
The standard ALICE event selection~\cite{Aamodt:2010pb} and centrality
estimator based on the V0--amplitude~\cite{Abelev:2013qoq} are used in
this analysis.  The event selection consists of: exclusion of
background events using the timing information from the ZDC and V0
detectors; verification of the trigger conditions; and a reconstructed
position of the collision.  As discussed
elsewhere~\cite{Abelev:2013qoq}, the \centRange{90}{100} centrality
class has substantial contributions from QED processes and is
therefore not included in the results presented here.

\belowpdfbookmark{Central tracklet measurements}{data:central}%
The measurement of the charged-particle pseudorapidity density at
mid-rapidity ($|\eta|<2$) is obtained from a tracklet analysis using
the two layers of the SPD.  The analysis method used is identical to
what has previously been
presented~\cite{Abbas:2013bpa,Aamodt:2010cz,Adam:2015ptt}.  Note that
no attempt is made to correct for known deficiencies, such as
deviations in the number of strange particles or transverse momentum
($\pT$) distributions compared to experimental
measurements~\cite{Abelev:2013ila,Abelev:2013xaa,Abelev:2013qoq}, in
the event generators used to obtain the corrections from simulations
(e.g., HIJING). 
It is found, through simulation studies, that tracklet reconstruction
first and foremost depends on the local hit density and only weakly on
particle mix and transverse momentum.
For example, the deficit of strange particles in the event generator
effects the result by less than \per{2}.  Since the event generators
generally, after detector simulation,  produce a local hit density
that  is consistent with what is observed in data, we observe a
correspondence between the tracklet samples of both simulations and
data. On the other hand, changing the number of tracklets
corresponding to strange particles a postiori to match the measured
relative yields dramatically biases the simulated tracklet sample away
from the measured, thus entailing systematic uncertainties that are
beyond the effect of the known event  generator deficiencies, and as
such do not improve the accuracy of the measurements.
Instead, variations on the event generators are used to
estimate the systematic uncertainties as detailed elsewhere
\cite{Abbas:2013bpa,Aamodt:2010cz,Adam:2015ptt}.
 
\belowpdfbookmark{Forward measurements}{data:forward}%
In the forward regions ($-3.5<\eta<-1.8$ and $1.8<\eta<5$), the
measurement is provided by the analysis of the deposited energy signal
in the FMD.  The analysis method used is identical to what has
previously been presented~\cite{Adam:2015kda,Abbas:2013bpa}: a
statistical approach to calculate the inclusive number of charged
particles; and a data-driven correction --- derived from previous
satellite--main collisions --- to remove the large background from
secondary particles.



\section{Systematic uncertainties}
\label{sec:sysuncer}

\belowpdfbookmark{Midrapidity}{sysuncer:central}%
For the measurements at mid-rapidity the sources and dependencies of
the systematic uncertainties are detailed elsewhere
\cite{Aamodt:2010cz,Adam:2015kda,Adam:2015ptt}.  The magnitude of the
systematic uncertainties is unchanged with respect to previous
results, and amounts to $\per{2.6}$ at $\eta=0$ and $\per{2.9}$ at
$\eta=2$, most of which is correlated over $|\eta|<2$, and largely
independent of centrality. 
 
\belowpdfbookmark{Forward rapidities}{sysuncer:forward}%
The systematic uncertainty on the forward analysis is evaluated using
the same technique as for previous results \cite{Adam:2015kda}. We
find that the uncertainty is uncorrelated across $\eta$ an that it
amounts to $\per{6.9}$ for $\eta>3.5$ and $\per{6.4}$ elsewhere within
the forward regions.

\belowpdfbookmark{Centrality}{sysuncer:centrality}%
The systematic uncertainty on $\ndndeta$ due to the centrality class
definition is estimated as $\per{0.6}$ for the most central and
$\per{9.5}$ for the most peripheral class \cite{Adam:2015ptt}. The
uncertainty is estimated by using alternative centrality definitions
based on SPD hit multiplicities and by varying the fraction of the
visible hadronic cross-section.  The \centRange{80}{90} centrality
class has some residual contamination from electromagnetic processes
detailed elsewhere~\cite{Abelev:2013qoq}, which gives rise to a
\per{4} additional systematic uncertainty on the measurements.

In summary, the total systematic uncertainty varies from $\per{2.6}$
at mid-rapidity in the most central collisions to $\per{12.4}$ at the
very forward rapidities for the most peripheral collisions.



\section{Results} 
\label{sec:results}

\begin{figure}[t!]
  \centering
  \result{dNdeta_05023}
\end{figure}

\belowpdfbookmark{dNch/deta}{results:dNdeta_05023}%
\figref*{fig:dNdeta_05023} presents the charged-particle
pseudorapidity density as a function of pseudorapidity for ten
centrality classes.  The measurements from the SPD and FMD are
combined in regions of overlap ($1.8<|\eta|<2$) between the two
detectors by taking the weighted average using the non-shared
uncertainties as weights.  Finally, based on the symmetry of the
collision system, the result is symmetrised around $\eta=0$, and
extended into the non-measured region $-5<\eta<-3.5$ by reflecting the
$3.5<\eta<5$ values around $\eta=0$.  Complementing result previously
reported at mid-rapidity \cite{Adam:2015ptt}, we find
$\dndeta{|_{|\eta|<0.5}}=17.52\pm0.05(\text{stat})\pm1.84(\text{sys})$
and $N_{\mathrm{part}}=7.3\pm0.1$ in the \centRange{80}{90} centrality
class.

\belowpdfbookmark{Total Nch vs. Npart}{results:TotalNch_05023}%
The measured distributions are fitted with four functions
$f_{\mathrm{GG}}$, $f_{\mathrm{P}}$, $f_{\mathrm{T}}$, and
$f_{\mathrm{B}}$ \cite{Adam:2015kda}, which are the difference of two
Gaussian distributions centred at $\eta=0$; a parametrisation proposed
by PHOBOS~\cite{PhysRevC.83.024913}; a trapezoidal form; and a plateau
connected to Gaussian tails, respectively.  To extract the total
number of charged particles, we calculate the integral and uncertainty
from the data in the measured region and use the integrals of the
fitted functions in the unmeasured regions up to the beam rapidity
$\pm y_{\text{beam}}=\pm 8.6$.  As for the previous measurements at
$\usNN{PbPb}{5023}$, the central value in the unmeasured regions
($-8.6<\eta<-3.5$ and $5<\eta<8.6$) is taken from the fit of the
function $f_{\mathrm{T}}$, while the uncertainty is evaluated as the
largest difference between the fitted functions scaled by
$1/\sqrt{3}$~\cite{Abbas:2013bpa,Adam:2015kda}. The total
charged-particle multiplicity is shown in \figref{fig:TotalNch_05023}
versus the mean number of participating nucleons
($\langle N_{\mathrm{part}}\rangle$) estimated from a Glauber
calculation~\cite{Abelev:2013qoq,Adam:2015ptt}.  After removing
correlated systematic uncertainties, we observe an increase in the
total number of charged particles of $\per{(27\pm4)}$ with respect to
the measurements at $\usNN{PbPb}{2760}$~\cite{Adam:2015kda} for all
centrality classes.  The line shown in \figref{fig:TotalNch_05023}
corresponds to a fit of a function inspired by
factorisation~\cite{PhysRevC.83.024913}.  The function illustrates
scaling by number of participant pairs, with a small perturbation
proportional to the cubic root of the number of participants.  As the
number of nucleon-nucleon collisions ($N_{\mathrm{coll}}$) scales
roughly like the square of the number of participants
$N_{\mathrm{coll}}\approx N_{\mathrm{part}}^2$~\cite{pubNote5023}, we
see no indication of scaling by number of nucleon--nucleon collisions.
The observed total $N_{\mathrm{ch}}$ dependence on
$\langle N_{\mathrm{part}}\rangle$ provides no evidence of any
significant increase in the number of hard scatterings between the
participating nucleons and partons.

\begin{figure}[t!]
  \centering
  \result[.92\linewidth]{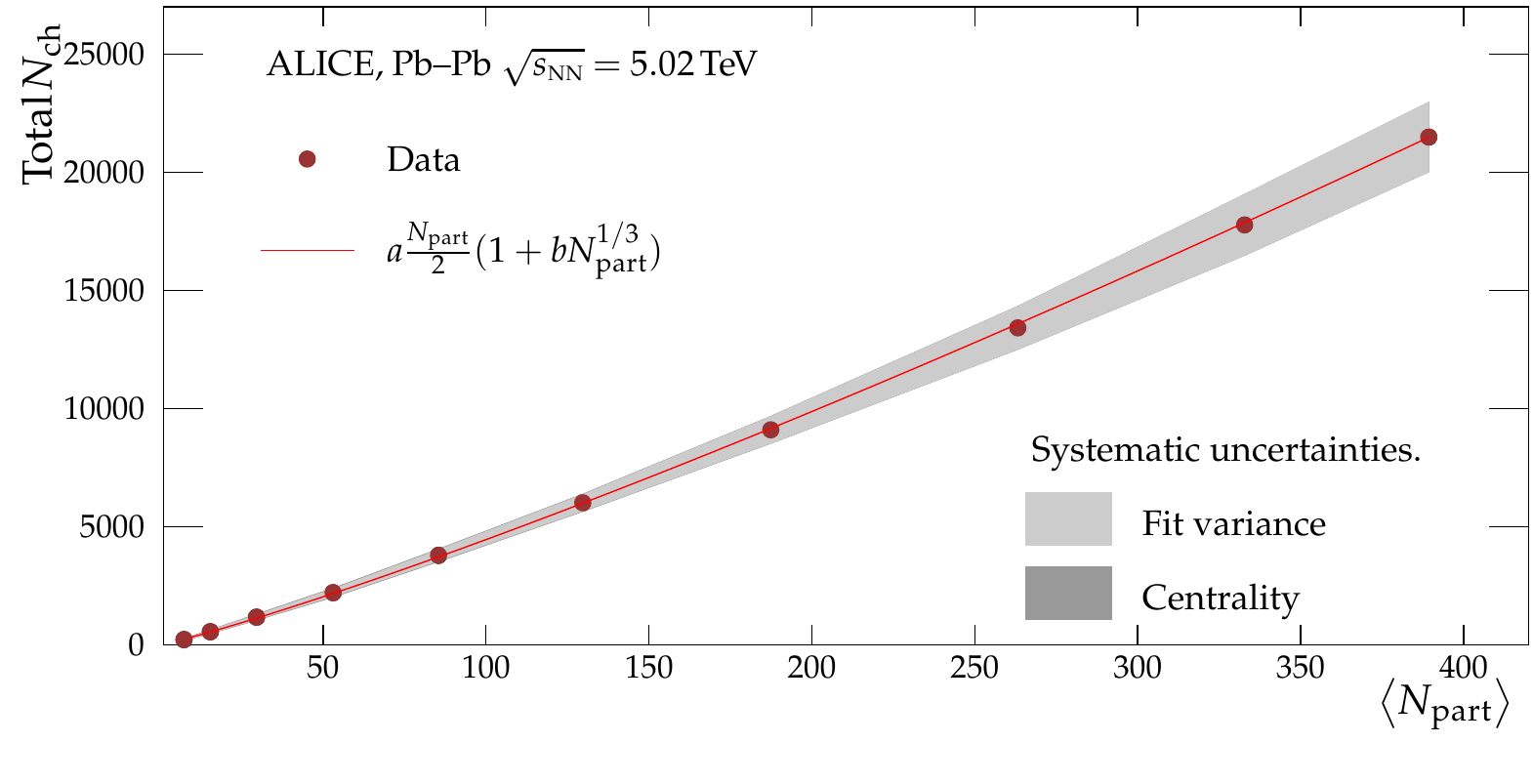}
\end{figure}
\begin{figure}[t!]
  \centering
  \result{Models1_05023}
\end{figure}

\belowpdfbookmark{Models}{results:Models1_05023}%
In \figref{fig:Models1_05023}, we compare the charged-particle
pseudorapidity density for the \centRange{0}{5} most central
collisions to three models: HIJING~\cite{Wang:1991hta};
EPOS--LHC~\cite{Pierog:2013ria}; and
KLN~\cite{Kharzeev:2004if,Dumitru:2011wq}, also for the
\centRange{0}{5} most central, except for KLN which is shown for the
\centRange{0}{6} centrality class.  Two versions of HIJING are used:
version~1.383, with jet quenching disabled, shadowing enabled, and a
hard $p_{_{\mathrm{T}}}$ cut-off of \unit[2.3]{Ge\kern-.2exV}; and the
newer version~2.1~\cite{Deng:2010mv}.  Both are two-component models
with a soft and hard sector defined by a $p_{_{\mathrm{T}}}$ cut-off
separating the two. In the 2.1 implementation, HIJING uses an upgraded
parametrisation of the nuclear parton distribution functions. This
results in a larger cross section for soft processes and a smaller
cross section for jet production.  The KLN~model is based on
Colour-Glass-Condensate initial conditions, while EPOS-LHC uses
so-called parton-ladders which hadronise in a medium.  While none of
the three models describe the measured charged-particle pseudorapidity
density over the full pseudorapidity range, we observe some
differences: HIJING~1.383 over-predicts the charged-particle
production especially away from $\eta\approx0$; EPOS--LHC and
HIJING~2.1 consistently under-predict the charge-particle production;
whereas KLN, EPOS--LHC, and HIJING~2.1 give a shape reasonably close
to the observed distribution.  
Not shown in \figref{fig:Models1_05023}, for both HIJING~1.383 and
EPOS--LHC, these observations hold over all centrality classes i.e.,
HIJING~1.383 consistently produces far too many particles away from
mid-rapidity and EPOS--LHC consistently under-predicts the
charged-particle yield over the full $\eta$ range.  These trends
become increasingly more pronounced for more peripheral collisions.

\begin{figure}[t!]
  \centering
  \result{TotalNchSys_05023}
\end{figure}

\belowpdfbookmark{Total Nch vs. sqrt(sNN)}{results:TotalNchSys_05023}%
\figref*{fig:TotalNchSys_05023} shows the total number of charged
particles produced in the most central heavy-ion collisions as a
function of the collision energy, ranging from $\usNN{PbPb}{2.6}$ to
$\TeV[5.02]$~\cite{Abbas:2013bpa}.  The dotted, dashed, and full-drawn
lines in the figure represent extrapolations from lower energy results
to the current top LHC energy of $\usNN{PbPb}{5023}$.  None of these
predictions fully describe the data.  A refit of the simple model of a
logarithmic-dampened power-law in the square collision energy ($s$)
including from the lowest to the highest energy results, shown as the
dash-dotted line, does accurately describe the total number of charged
particles at all available energies.

\begin{figure}[t!]
  \centering
  \result{dNdy_05023}
\end{figure}

\belowpdfbookmark{dN/dy}{results:dNdy_05023}%
We can calculate the Jacobian transform from $\eta$ to rapidity $y$ by
assuming the same transverse momentum distribution of
(anti\nolinebreak -)\nolinebreak protons, and charged kaons and pions,
and the same particle ratios in Pb--Pb collisions at
$\usNN{PbPb}{5023}$ as in $\usNN{PbPb}{2760}$.
The result is presented in \figref{fig:dNdy_05023} for the
\centRange{0}{5} most central collisions.  The effect on the Jacobian
from the change of $p_{_{\mathrm{T}}}$ spectra and particle ratios
when increasing the collision energy by almost a factor two is
evaluated using the EPOS--LHC model~\cite{Pierog:2013ria}.  It is
found, that the effect is at most $\unit[3]{\mbox{\textperthousand}}$
on both $\dndy$ and $y$ --- much smaller than the systematic
uncertainty and $\eta$ resolution of the analysis.
\figref*{fig:dNdy_05023} also shows the expected charged-particle
rapidity densities from the Landau-Carruthers \cite{Carruthers:1973rw}
and Landau-Wong \cite{Wong:2008ex} models, both assuming Landau
hydrodynamics i.e., based on a reaction scenario with full stopping of
the reaction partners and a subsequent thermodynamic evolution.  The
measurements, however, are seen to be consistent with a Gaussian
distribution with a width of $4.12\pm0.10$, much wider than the width
expected from the two models.  A best parameter fit of the sum of two
Gaussian distributions with means symmetric around $y=0$, is
indistinguishable from the single Gaussian case.

\begin{figure}[t!]
  \centering
  \bgroup   
  \def\myMarkSize{4pt}
  \def\myLabelScale{1.2}
  \def\myTitleScale{1.6}
  \def\myScale{1.6}
  \result{ScaledSigma_05023}
  \egroup
\end{figure}

\belowpdfbookmark{Scaled sigma}{results:ScaledSigma_05023}%
In the top part of \figref{fig:ScaledSigma_05023} we compare the
widths of the charged-particle or -pion rapidity density distribution
extracted from measurements to the expected width
$\sigma^2_{\text{L-C}}=\log(\usNN{PbPb}{}/2m_{p})$ from
Landau-Carruthers, where $m_p$ is the proton mass, at collision
energies ranging from $\GeV[2.6]{}$ up to $\TeV[5.02]{}$.  An increase
of $\approx\per{7}$ of $\sigma_{\dnxdy[x]{}}/\sigma_{\text{L-C}}$ is
seen from the $\usNN{PbPb}{2760}$ ALICE measurements
\cite{Abbas:2013bpa}. The full evolution is consistent with an almost
linear rise as a function of $\log{\usNN{PbPb}{}}$ from the top SPS
energy at $\usNN{PbPb}{17.3}$.  It can be shown~\cite{Mohanty:2003va}
that the width of the rapidity-density distribution in Landau
hydrodynamics scales as $\sigma_{\dnxdy[x]{}}\propto 1/(1-c_s^2)$,
where $c_s$ is the speed of sound in the matter. The lifetime of the
system scales inversely with $c_s$, and given that the measured width
is larger than the predicted by Landau hydrodynamics, it is an
indication that, given the considerations above, the lifetime is
shorter than suggested. 

In the bottom part of \figref{fig:ScaledSigma_05023} we compare the
width of the $\dndy$ distribution to the available rapidity range
($2y_{\mathrm{beam}}$).  We observe no dependence of this ratio from
$\usNN{PbPb}{17.3}$ and upward, indicating that the available
phase-space constrains the width of that distribution.  The
charged-hadron measurements at RHIC (crosses) from the
BRAHMS~\cite{Bearden:2001qq} and PHOBOS~\cite{Back:2002wb}
measurements of $\dndeta$ are converted to $\dndy$ using the same
method as applied to the ALICE data.  Previously, charged-pion
measurements from BRAHMS have been reported~\cite{Bearden:2004yx}.
These data are not included because a re-evaluation using RHIC Run--4
Au--Au data has not been finalised \cite{Videbaek:2009zy}.

From the observed $s^p$ scaling of the charged-particle pseudorapidity
density at mid-rapidity \cite{Adam:2015ptt} we expect a $\per{20}$
increase over $\usNN{PbPb}{2760}$ in the level of
$\dndeta{|_{|\eta|<0.5}}$ and from the extracted width of $\dndy$ we
observe an additional $\per{7}$, consistent with the increase of 
$\per{27}$ over $\usNN{PbPb}{2760}$ in the total number of charged
particles produced in $\usNN{PbPb}{5023}$ collisions.



\section{Conclusions}
\label{sec:concl}

The charged-particle pseudorapidity density is measured in Pb--Pb
collisions at $\usNN{PbPb}{5023}$ over the psuedorapidity range
$-3.5<\eta<5$.  The total number of charged particles produced is
determined owing to the large pseudorapidity acceptance of ALICE.  The
latter increases by two orders of magnitude from the most peripheral
to the most central collisions and scales approximately with the
number of participating nucleons.  The increase in the total number of
charged particles relative to $\usNN{PbPb}{2760}$ is estimated to be
$\per{(27\pm4)}$.  The charged-particle rapidity density for the most
central collisions is extracted, and the width of that distribution is
compared to predictions from the Landau-Carruthers and Landau-Wong
hydrodynamic models.
It is found that the measured charged-particle rapidity density
becomes increasingly wider as a function of collision energy than
predicted by Landau hydrodynamics. The width of the charged-particle
rapidity density is seen to scale with the beam rapidity, which
implies that the available phase space determines the longitudinal
extend of the charged-particle production. The phase space dominance
starts at the top SPS energy and persist for two orders of magnitude
up to the top LHC energy.

\currentpdfbookmark{Acknowledgements}{ack}%
\section*{Acknowledgements}
\label{app:acknowledgements}

The ALICE Collaboration would like to thank all its engineers and technicians for their invaluable contributions to the construction of the experiment and the CERN accelerator teams for the outstanding performance of the LHC complex.
The ALICE Collaboration gratefully acknowledges the resources and support provided by all Grid centres and the Worldwide LHC Computing Grid (WLCG) collaboration.
The ALICE Collaboration acknowledges the following funding agencies for their support in building and running the ALICE detector:
A. I. Alikhanyan National Science Laboratory (Yerevan Physics Institute) Foundation (ANSL), State Committee of Science and World Federation of Scientists (WFS), Armenia;
Austrian Academy of Sciences and Nationalstiftung f\"{u}r Forschung, Technologie und Entwicklung, Austria;
Conselho Nacional de Desenvolvimento Cient\'{\i}fico e Tecnol\'{o}gico (CNPq), Universidade Federal do Rio Grande do Sul (UFRGS), Financiadora de Estudos e Projetos (Finep) and Funda\c{c}\~{a}o de Amparo \`{a} Pesquisa do Estado de S\~{a}o Paulo (FAPESP), Brazil;
Ministry of Science \& Technology of China (MSTC), National Natural Science Foundation of China (NSFC) and Ministry of Education of China (MOEC) , China;
Ministry of Science, Education and Sport and Croatian Science Foundation, Croatia;
Ministry of Education, Youth and Sports of the Czech Republic, Czech Republic;
The Danish Council for Independent Research | Natural Sciences, the Carlsberg Foundation and Danish National Research Foundation (DNRF), Denmark;
Helsinki Institute of Physics (HIP), Finland;
Commissariat \`{a} l'Energie Atomique (CEA) and Institut National de Physique Nucl\'{e}aire et de Physique des Particules (IN2P3) and Centre National de la Recherche Scientifique (CNRS), France;
Bundesministerium f\"{u}r Bildung, Wissenschaft, Forschung und Technologie (BMBF) and GSI Helmholtzzentrum f\"{u}r Schwerionenforschung GmbH, Germany;
Ministry of Education, Research and Religious Affairs, Greece;
National Research, Development and Innovation Office, Hungary;
Department of Atomic Energy Government of India (DAE) and Council of Scientific and Industrial Research (CSIR), New Delhi, India;
Indonesian Institute of Science, Indonesia;
Centro Fermi - Museo Storico della Fisica e Centro Studi e Ricerche Enrico Fermi and Istituto Nazionale di Fisica Nucleare (INFN), Italy;
Institute for Innovative Science and Technology , Nagasaki Institute of Applied Science (IIST), Japan Society for the Promotion of Science (JSPS) KAKENHI and Japanese Ministry of Education, Culture, Sports, Science and Technology (MEXT), Japan;
Consejo Nacional de Ciencia (CONACYT) y Tecnolog\'{i}a, through Fondo de Cooperaci\'{o}n Internacional en Ciencia y Tecnolog\'{i}a (FONCICYT) and Direcci\'{o}n General de Asuntos del Personal Academico (DGAPA), Mexico;
Nationaal instituut voor subatomaire fysica (Nikhef), Netherlands;
The Research Council of Norway, Norway;
Commission on Science and Technology for Sustainable Development in the South (COMSATS), Pakistan;
Pontificia Universidad Cat\'{o}lica del Per\'{u}, Peru;
Ministry of Science and Higher Education and National Science Centre, Poland;
Korea Institute of Science and Technology Information and National Research Foundation of Korea (NRF), Republic of Korea;
Ministry of Education and Scientific Research, Institute of Atomic Physics and Romanian National Agency for Science, Technology and Innovation, Romania;
Joint Institute for Nuclear Research (JINR), Ministry of Education and Science of the Russian Federation and National Research Centre Kurchatov Institute, Russia;
Ministry of Education, Science, Research and Sport of the Slovak Republic, Slovakia;
National Research Foundation of South Africa, South Africa;
Centro de Aplicaciones Tecnol\'{o}gicas y Desarrollo Nuclear (CEADEN), Cubaenerg\'{\i}a, Cuba, Ministerio de Ciencia e Innovacion and Centro de Investigaciones Energ\'{e}ticas, Medioambientales y Tecnol\'{o}gicas (CIEMAT), Spain;
Swedish Research Council (VR) and Knut \& Alice Wallenberg Foundation (KAW), Sweden;
European Organization for Nuclear Research, Switzerland;
National Science and Technology Development Agency (NSDTA), Suranaree University of Technology (SUT) and Office of the Higher Education Commission under NRU project of Thailand, Thailand;
Turkish Atomic Energy Agency (TAEK), Turkey;
National Academy of  Sciences of Ukraine, Ukraine;
Science and Technology Facilities Council (STFC), United Kingdom;
National Science Foundation of the United States of America (NSF) and United States Department of Energy, Office of Nuclear Physics (DOE NP), United States of America.
\currentpdfbookmark{References}{bib}%
\bibliography{bib}

\clearpage
\appendix 
\section{The ALICE Collaboration}
\label{app:collab}



\begingroup
\small
\begin{flushleft}
J.~Adam$^\textrm{\scriptsize 38}$,
D.~Adamov\'{a}$^\textrm{\scriptsize 86}$,
M.M.~Aggarwal$^\textrm{\scriptsize 90}$,
G.~Aglieri Rinella$^\textrm{\scriptsize 34}$,
M.~Agnello$^\textrm{\scriptsize 30}$\textsuperscript{,}$^\textrm{\scriptsize 112}$,
N.~Agrawal$^\textrm{\scriptsize 47}$,
Z.~Ahammed$^\textrm{\scriptsize 137}$,
S.~Ahmad$^\textrm{\scriptsize 17}$,
S.U.~Ahn$^\textrm{\scriptsize 69}$,
S.~Aiola$^\textrm{\scriptsize 141}$
% ...
\renewcommand\labelenumi{\textsuperscript{\theenumi}~}

\section*{Affiliation notes}
\renewcommand\theenumi{\roman{enumi}}
\begin{Authlist}
\item \Adef{0}Deceased
\item \Adef{idp1792368}{Also at: Georgia State University, Atlanta, Georgia, United States}
\item \Adef{idp3223552}{Also at: Also at Department of Applied Physics, Aligarh Muslim University, Aligarh, India}
\item \Adef{idp3995136}{Also at: M.V. Lomonosov Moscow State University, D.V. Skobeltsyn Institute of Nuclear, Physics, Moscow, Russia}
\end{Authlist}

\section*{Collaboration Institutes}
\renewcommand\theenumi{\arabic{enumi}~}

$^{1}$A.I. Alikhanyan National Science Laboratory (Yerevan Physics Institute) Foundation, Yerevan, Armenia
\\
$^{2}$Benem\'{e}rita Universidad Aut\'{o}noma de Puebla, Puebla, Mexico
\\
$^{3}$Bogolyubov Institute for Theoretical Physics, Kiev, Ukraine
\\
$^{4}$Bose Institute, Department of Physics 
and Centre for Astroparticle Physics and Space Science (CAPSS), Kolkata, India
\\
$^{5}$Budker Institute for Nuclear Physics, Novosibirsk, Russia
\\
$^{6}$California Polytechnic State University, San Luis Obispo, California, United States
\\
$^{7}$Central China Normal University, Wuhan, China
\\
$^{8}$Centre de Calcul de l'IN2P3, Villeurbanne, Lyon, France
\\
$^{9}$Centro de Aplicaciones Tecnol\'{o}gicas y Desarrollo Nuclear (CEADEN), Havana, Cuba
\\
$^{10}$Centro de Investigaciones Energ\'{e}ticas Medioambientales y Tecnol\'{o}gicas (CIEMAT), Madrid, Spain
\\
$^{11}$Centro de Investigaci\'{o}n y de Estudios Avanzados (CINVESTAV), Mexico City and M\'{e}rida, Mexico
\\
$^{12}$Centro Fermi - Museo Storico della Fisica e Centro Studi e Ricerche ``Enrico Fermi', Rome, Italy
\\
$^{13}$Chicago State University, Chicago, Illinois, United States
\\
$^{14}$China Institute of Atomic Energy, Beijing, China
\\
$^{15}$COMSATS Institute of Information Technology (CIIT), Islamabad, Pakistan
\\
$^{16}$Departamento de F\'{\i}sica de Part\'{\i}culas and IGFAE, Universidad de Santiago de Compostela, Santiago de Compostela, Spain
\\
$^{17}$Department of Physics, Aligarh Muslim University, Aligarh, India
\\
$^{18}$Department of Physics, Ohio State University, Columbus, Ohio, United States
\\
$^{19}$Department of Physics, Sejong University, Seoul, South Korea
\\
$^{20}$Department of Physics, University of Oslo, Oslo, Norway
\\
$^{21}$Department of Physics and Technology, University of Bergen, Bergen, Norway
\\
$^{22}$Dipartimento di Fisica dell'Universit\`{a} 'La Sapienza'
and Sezione INFN, Rome, Italy
\\
$^{23}$Dipartimento di Fisica dell'Universit\`{a}
and Sezione INFN, Cagliari, Italy
\\
$^{24}$Dipartimento di Fisica dell'Universit\`{a}
and Sezione INFN, Trieste, Italy
\\
$^{25}$Dipartimento di Fisica dell'Universit\`{a}
and Sezione INFN, Turin, Italy
\\
$^{26}$Dipartimento di Fisica e Astronomia dell'Universit\`{a}
and Sezione INFN, Bologna, Italy
\\
$^{27}$Dipartimento di Fisica e Astronomia dell'Universit\`{a}
and Sezione INFN, Catania, Italy
\\
$^{28}$Dipartimento di Fisica e Astronomia dell'Universit\`{a}
and Sezione INFN, Padova, Italy
\\
$^{29}$Dipartimento di Fisica `E.R.~Caianiello' dell'Universit\`{a}
and Gruppo Collegato INFN, Salerno, Italy
\\
$^{30}$Dipartimento DISAT del Politecnico and Sezione INFN, Turin, Italy
\\
$^{31}$Dipartimento di Scienze e Innovazione Tecnologica dell'Universit\`{a} del Piemonte Orientale and INFN Sezione di Torino, Alessandria, Italy
\\
$^{32}$Dipartimento Interateneo di Fisica `M.~Merlin'
and Sezione INFN, Bari, Italy
\\
$^{33}$Division of Experimental High Energy Physics, University of Lund, Lund, Sweden
\\
$^{34}$European Organization for Nuclear Research (CERN), Geneva, Switzerland
\\
$^{35}$Excellence Cluster Universe, Technische Universit\"{a}t M\"{u}nchen, Munich, Germany
\\
$^{36}$Faculty of Engineering, Bergen University College, Bergen, Norway
\\
$^{37}$Faculty of Mathematics, Physics and Informatics, Comenius University, Bratislava, Slovakia
\\
$^{38}$Faculty of Nuclear Sciences and Physical Engineering, Czech Technical University in Prague, Prague, Czech Republic
\\
$^{39}$Faculty of Science, P.J.~\v{S}af\'{a}rik University, Ko\v{s}ice, Slovakia
\\
$^{40}$Faculty of Technology, Buskerud and Vestfold University College, Tonsberg, Norway
\\
$^{41}$Frankfurt Institute for Advanced Studies, Johann Wolfgang Goethe-Universit\"{a}t Frankfurt, Frankfurt, Germany
\\
$^{42}$Gangneung-Wonju National University, Gangneung, South Korea
\\
$^{43}$Gauhati University, Department of Physics, Guwahati, India
\\
$^{44}$Helmholtz-Institut f\"{u}r Strahlen- und Kernphysik, Rheinische Friedrich-Wilhelms-Universit\"{a}t Bonn, Bonn, Germany
\\
$^{45}$Helsinki Institute of Physics (HIP), Helsinki, Finland
\\
$^{46}$Hiroshima University, Hiroshima, Japan
\\
$^{47}$Indian Institute of Technology Bombay (IIT), Mumbai, India
\\
$^{48}$Indian Institute of Technology Indore, Indore, India
\\
$^{49}$Indonesian Institute of Sciences, Jakarta, Indonesia
\\
$^{50}$Inha University, Incheon, South Korea
\\
$^{51}$Institut de Physique Nucl\'eaire d'Orsay (IPNO), Universit\'e Paris-Sud, CNRS-IN2P3, Orsay, France
\\
$^{52}$Institute for Nuclear Research, Academy of Sciences, Moscow, Russia
\\
$^{53}$Institute for Subatomic Physics of Utrecht University, Utrecht, Netherlands
\\
$^{54}$Institute for Theoretical and Experimental Physics, Moscow, Russia
\\
$^{55}$Institute of Experimental Physics, Slovak Academy of Sciences, Ko\v{s}ice, Slovakia
\\
$^{56}$Institute of Physics, Academy of Sciences of the Czech Republic, Prague, Czech Republic
\\
$^{57}$Institute of Physics, Bhubaneswar, India
\\
$^{58}$Institute of Space Science (ISS), Bucharest, Romania
\\
$^{59}$Institut f\"{u}r Informatik, Johann Wolfgang Goethe-Universit\"{a}t Frankfurt, Frankfurt, Germany
\\
$^{60}$Institut f\"{u}r Kernphysik, Johann Wolfgang Goethe-Universit\"{a}t Frankfurt, Frankfurt, Germany
\\
$^{61}$Institut f\"{u}r Kernphysik, Westf\"{a}lische Wilhelms-Universit\"{a}t M\"{u}nster, M\"{u}nster, Germany
\\
$^{62}$Instituto de Ciencias Nucleares, Universidad Nacional Aut\'{o}noma de M\'{e}xico, Mexico City, Mexico
\\
$^{63}$Instituto de F\'{i}sica, Universidade Federal do Rio Grande do Sul (UFRGS), Porto Alegre, Brazil
\\
$^{64}$Instituto de F\'{\i}sica, Universidad Nacional Aut\'{o}noma de M\'{e}xico, Mexico City, Mexico
\\
$^{65}$IRFU, CEA, Universit\'{e} Paris-Saclay, F-91191 Gif-sur-Yvette, France, Saclay, France
\\
$^{66}$iThemba LABS, National Research Foundation, Somerset West, South Africa
\\
$^{67}$Joint Institute for Nuclear Research (JINR), Dubna, Russia
\\
$^{68}$Konkuk University, Seoul, South Korea
\\
$^{69}$Korea Institute of Science and Technology Information, Daejeon, South Korea
\\
$^{70}$KTO Karatay University, Konya, Turkey
\\
$^{71}$Laboratoire de Physique Corpusculaire (LPC), Clermont Universit\'{e}, Universit\'{e} Blaise Pascal, CNRS--IN2P3, Clermont-Ferrand, France
\\
$^{72}$Laboratoire de Physique Subatomique et de Cosmologie, Universit\'{e} Grenoble-Alpes, CNRS-IN2P3, Grenoble, France
\\
$^{73}$Laboratori Nazionali di Frascati, INFN, Frascati, Italy
\\
$^{74}$Laboratori Nazionali di Legnaro, INFN, Legnaro, Italy
\\
$^{75}$Lawrence Berkeley National Laboratory, Berkeley, California, United States
\\
$^{76}$Moscow Engineering Physics Institute, Moscow, Russia
\\
$^{77}$Nagasaki Institute of Applied Science, Nagasaki, Japan
\\
$^{78}$National and Kapodistrian University of Athens, Physics Department, Athens, Greece, Athens, Greece
\\
$^{79}$National Centre for Nuclear Studies, Warsaw, Poland
\\
$^{80}$National Institute for Physics and Nuclear Engineering, Bucharest, Romania
\\
$^{81}$National Institute of Science Education and Research, Bhubaneswar, India
\\
$^{82}$National Research Centre Kurchatov Institute, Moscow, Russia
\\
$^{83}$Niels Bohr Institute, University of Copenhagen, Copenhagen, Denmark
\\
$^{84}$Nikhef, Nationaal instituut voor subatomaire fysica, Amsterdam, Netherlands
\\
$^{85}$Nuclear Physics Group, STFC Daresbury Laboratory, Daresbury, United Kingdom
\\
$^{86}$Nuclear Physics Institute, Academy of Sciences of the Czech Republic, \v{R}e\v{z} u Prahy, Czech Republic
\\
$^{87}$Oak Ridge National Laboratory, Oak Ridge, Tennessee, United States
\\
$^{88}$Petersburg Nuclear Physics Institute, Gatchina, Russia
\\
$^{89}$Physics Department, Creighton University, Omaha, Nebraska, United States
\\
$^{90}$Physics Department, Panjab University, Chandigarh, India
\\
$^{91}$Physics Department, University of Cape Town, Cape Town, South Africa
\\
$^{92}$Physics Department, University of Jammu, Jammu, India
\\
$^{93}$Physics Department, University of Rajasthan, Jaipur, India
\\
$^{94}$Physikalisches Institut, Eberhard Karls Universit\"{a}t T\"{u}bingen, T\"{u}bingen, Germany
\\
$^{95}$Physikalisches Institut, Ruprecht-Karls-Universit\"{a}t Heidelberg, Heidelberg, Germany
\\
$^{96}$Physik Department, Technische Universit\"{a}t M\"{u}nchen, Munich, Germany
\\
$^{97}$Purdue University, West Lafayette, Indiana, United States
\\
$^{98}$Pusan National University, Pusan, South Korea
\\
$^{99}$Research Division and ExtreMe Matter Institute EMMI, GSI Helmholtzzentrum f\"ur Schwerionenforschung GmbH, Darmstadt, Germany
\\
$^{100}$Rudjer Bo\v{s}kovi\'{c} Institute, Zagreb, Croatia
\\
$^{101}$Russian Federal Nuclear Center (VNIIEF), Sarov, Russia
\\
$^{102}$Saha Institute of Nuclear Physics, Kolkata, India
\\
$^{103}$School of Physics and Astronomy, University of Birmingham, Birmingham, United Kingdom
\\
$^{104}$Secci\'{o}n F\'{\i}sica, Departamento de Ciencias, Pontificia Universidad Cat\'{o}lica del Per\'{u}, Lima, Peru
\\
$^{105}$Sezione INFN, Bari, Italy
\\
$^{106}$Sezione INFN, Bologna, Italy
\\
$^{107}$Sezione INFN, Cagliari, Italy
\\
$^{108}$Sezione INFN, Catania, Italy
\\
$^{109}$Sezione INFN, Padova, Italy
\\
$^{110}$Sezione INFN, Rome, Italy
\\
$^{111}$Sezione INFN, Trieste, Italy
\\
$^{112}$Sezione INFN, Turin, Italy
\\
$^{113}$SSC IHEP of NRC Kurchatov institute, Protvino, Russia
\\
$^{114}$Stefan Meyer Institut f\"{u}r Subatomare Physik (SMI), Vienna, Austria
\\
$^{115}$SUBATECH, Ecole des Mines de Nantes, Universit\'{e} de Nantes, CNRS-IN2P3, Nantes, France
\\
$^{116}$Suranaree University of Technology, Nakhon Ratchasima, Thailand
\\
$^{117}$Technical University of Ko\v{s}ice, Ko\v{s}ice, Slovakia
\\
$^{118}$Technical University of Split FESB, Split, Croatia
\\
$^{119}$The Henryk Niewodniczanski Institute of Nuclear Physics, Polish Academy of Sciences, Cracow, Poland
\\
$^{120}$The University of Texas at Austin, Physics Department, Austin, Texas, United States
\\
$^{121}$Universidad Aut\'{o}noma de Sinaloa, Culiac\'{a}n, Mexico
\\
$^{122}$Universidade de S\~{a}o Paulo (USP), S\~{a}o Paulo, Brazil
\\
$^{123}$Universidade Estadual de Campinas (UNICAMP), Campinas, Brazil
\\
$^{124}$Universidade Federal do ABC, Santo Andre, Brazil
\\
$^{125}$University of Houston, Houston, Texas, United States
\\
$^{126}$University of Jyv\"{a}skyl\"{a}, Jyv\"{a}skyl\"{a}, Finland
\\
$^{127}$University of Liverpool, Liverpool, United Kingdom
\\
$^{128}$University of Tennessee, Knoxville, Tennessee, United States
\\
$^{129}$University of the Witwatersrand, Johannesburg, South Africa
\\
$^{130}$University of Tokyo, Tokyo, Japan
\\
$^{131}$University of Tsukuba, Tsukuba, Japan
\\
$^{132}$University of Zagreb, Zagreb, Croatia
\\
$^{133}$Universit\'{e} de Lyon, Universit\'{e} Lyon 1, CNRS/IN2P3, IPN-Lyon, Villeurbanne, Lyon, France
\\
$^{134}$Universit\'{e} de Strasbourg, CNRS, IPHC UMR 7178, F-67000 Strasbourg, France, Strasbourg, France
\\
$^{135}$Universit\`{a} di Brescia, Brescia, Italy
\\
$^{136}$V.~Fock Institute for Physics, St. Petersburg State University, St. Petersburg, Russia
\\
$^{137}$Variable Energy Cyclotron Centre, Kolkata, India
\\
$^{138}$Warsaw University of Technology, Warsaw, Poland
\\
$^{139}$Wayne State University, Detroit, Michigan, United States
\\
$^{140}$Wigner Research Centre for Physics, Hungarian Academy of Sciences, Budapest, Hungary
\\
$^{141}$Yale University, New Haven, Connecticut, United States
\\
$^{142}$Yonsei University, Seoul, South Korea
\\
$^{143}$Zentrum f\"{u}r Technologietransfer und Telekommunikation (ZTT), Fachhochschule Worms, Worms, Germany
\endgroup

\end{document}